\documentstyle{mn}

\newif\ifAMStwofonts
\newcommand{\etal}{{et al.}~}
\newcommand{\de}{\delta}
\newcommand{\te}{\theta}
\newcommand{\varte}{\vartheta}
\newcommand{\veps}{\varepsilon}
\newcommand{\Sig}{\Sigma}

\newcommand{\f}{\frac}
\newcommand{\s}{\sigma}
\newcommand{\bfpsi}{\bmath{\psi}}
\newcommand{\bfx}{\bmath{x}}

\newcommand{\bfk}{\bmath{k}}
\newcommand{\bfv}{\bmath{v}}
\newcommand{\bfq}{\bmath{q}}

\newcommand{\calO}{{\cal O}}

\newcommand{\calZ}{{\cal Z}}

\newcommand{\bc}{\begin{center}}
\newcommand{\be}{\begin{equation}}
\newcommand{\ee}{\end{equation}}
\newcommand{\ec}{\end{center}}
\newcommand{\lan}{\langle}
\newcommand{\ran}{\rangle}

\newcommand{\hmpc}{$h^{-1}\,{\rm Mpc}$}
\newcommand{\kms}{{${\rm km\; s^{-1}}$}}

\title[Density vs velocity-divergence relation in the Zel'dovich 
	approximation]{Cosmological density versus 
	velocity-divergence relation in the Zel'dovich approximation}

\author[M. J. Chodorowski]
     {Micha{\l} J.\ Chodorowski\\
      Copernicus Astronomical Center, Bartycka 18,
     00--716 Warsaw, Poland\\
          }

\begin{document}

\maketitle

\begin{abstract}
I derive a relation, both `forward' and `inverse', between the density
and the divergence of the peculiar velocity which results from the
Zel'dovich approximation. My calculations assume Gaussian initial
conditions. The forward relation expresses the density (strictly
speaking, the expectation value of the continuity density given the
velocity divergence) in terms of the velocity divergence, while the
inverse relation expresses the velocity divergence in terms of the
density. The predicted scatter in the relations is small, hence the
inverse relation is close to, though not identical with, a
mathematical inversion of the forward one. The forward relation is
equivalent to the well-known `standard' density--velocity relation in
the Zel'dovich approximation. The inverse relation, however, is
successfully derived for the first time and constitutes a potentially
interesting alternative to an inverse relation derived by Chodorowski
et al., based on third-order perturbation theory. Specifically, it may
better recover the peculiar velocity from the associated density
field, when smoothed over scales as small as a few megaparsecs.

\end{abstract}


\section{Introduction}
\label{sec:intro}
In the gravitational instability scenario for the formation of structure in
the Universe, the peculiar motions of galaxies are tightly related to the
large-scale mass distribution. The comparison between the density and the
velocity fields can serve as a test of the gravitational instability
hypothesis and as a method for estimating the cosmological parameter $\Omega$
(Dekel \etal 1993). In linear regime, the relation between the density and the
velocity fields is
\be
\de(\bfx) = - f(\Omega)^{-1} \bmath{\nabla} \cdot \bfv(\bfx) \,,
\label{eq:i1}
\ee
where $f(\Omega) \simeq \Omega^{0.6}$ and I express distances in
units of \kms. This equation is applicable only when the density
fluctuations are small compared to unity. However, sampling of
galaxies in current redshift surveys and random errors in peculiar
velocity catalogs enable reliable dynamical analysis with smoothing
scale of several \hmpc, where fluctuations slightly exceed the regime
of applicability of linear theory.

Relation~(\ref{eq:i1}) has been recently extended for the mildly non-linear
regime by Chodorowski \& {\L}okas~(1997; hereafter C{\L}). Let us define a
variable proportional to the velocity divergence,
\be
\varte \equiv - f(\Omega)^{-1} \bmath{\nabla} \cdot \bfv(\bfx) \,.
\label{eq:i2}
\ee 
C{\L} rigorously computed the mean $\de(\bfx)$ given $\varte(\bfx)$,
i.e., $\lan \de \ran|_{\varte}$, up to third order in (Eulerian)
perturbation theory (hereafter PT), assuming Gaussian initial
conditions. The resulting formula is

\be
\lan \de \ran|_{\varte} =
a_1 \varte + a_2 (\varte^2 - \veps_\varte^2) + a_3 \varte^3 \,,
\label{eq:i3}
\ee 
where $\veps_\varte^2$ is the variance of the field $\varte$. The
coefficients $a_i$ entering the above expansion are given by some
combinations of the joint moments of $\de$ and $\varte$ and were
explicitly calculated by C{\L}.  Gaussian initial conditions are also
assumed in the present paper.

Mildly non-linear relation between the density and the velocity
divergence is, in contrast to linear relation~(\ref{eq:i1}),
non-local. The local estimator of density~(\ref{eq:i3}) has thus a
non-zero variance. Therefore, to obtain an unbiased inverse estimator,
i.e., of the velocity divergence from the density, we cannot simply
invert expression~(\ref{eq:i3}). The inverse estimator was explicitly
constructed up to third order in PT by Chodorowski \etal (1998a;
hereafter C{\L}PN), who also computed the expected scatter in the
relation.

Having said that, it may seem difficult to understand why to
investigate the density versus velocity-divergence relation (hereafter
DVDR) in the Zel'dovich approximation (Zel'dovich 1970; hereafter
ZA). This approximation is first order in Lagrangian PT and therefore
provides only partial answers for higher-order perturbative
contributions to the density contrast and the velocity
divergence. Having solved the problem rigorously, why to resort to
approximate schemes again?

There are a few reasons for which the ZA is still worth
studying. Firstly, due to its simplicity, it is very popular and in
wide use. In particular, the density--velocity relation, resulting
from an Eulerian version of this approximation (Nusser \etal 1991), is
used in the {\sc potent} reconstruction of the mass density from
peculiar velocity data (Sigad \etal 1998). (Strictly speaking, Sigad
\etal use a formula {\em based\/} on the ZA, with the coefficients
slightly adjusted to best fit N-body results.) It is therefore
interesting to see how the ZA-based estimator of density relates to
perturbative formula~(\ref{eq:i3}).

Second, N-body simulations have shown that the ZA is apparently quite
successful in recovering the density from the corresponding velocity
field. An estimator of density resulting from the ZA (the {\em
continuity\/} density) happens to be merely slightly biased, even for
smoothing scales as small as a few \hmpc\ (Mancinelli \etal 1994,
Ganon \etal 1998), where PT is expected to break down. For a Gaussian
smoothing length of $5$ \hmpc, the rms fluctuation of a density field
is already close to unity and in the perturbative expansion for $\de$
and $\varte$, terms of all orders become comparable. Indeed, for
smoothing scales smaller than $5$ \hmpc, the value of the coefficient
$a_2$ estimated from N-body starts to deviate significantly from the
predicted value (Chodorowski \& Stompor 1998). This is not a problem
for formula~(\ref{eq:i3}), which is applicable to the mass density
reconstruction from peculiar velocities, a part of so-called
density--density comparisons.  These comparisons (e.g., {\it
IRAS}--{\sc potent}) currently employ Gaussian smoothing length of
$12$ \hmpc. For such a smoothing length, one may hope
formula~(\ref{eq:i3}) to be even better estimator of density than the
formula based on the ZA. Velocity--velocity comparisons, however,
employ smoothing lengths as small as $5$, or even $3$, \hmpc\ (e.g.,
Willick \& Strauss 1998). For such small scales, an inverted version
(i.e., an estimator of velocity from density) of the ZA may prove to
do better than a formula based on rigorous third-order PT. It is
therefore important to invert the ZA and to test its performance,
relative to a third-order perturbative formula, against N-body
simulations.

Finally, any galaxy density field is derived from a redshift survey,
i.e. given originally in the redshift space. To compare the galaxy
density field with the real-space mass density field inferred from
peculiar velocity data, the galaxy field must be first reconstructed
in the real space. The redshifts of galaxies differ from the true
distances by the peculiar velocities, induced themselves by the
fluctuations in the density field. Hence, the real-space galaxy
density reconstruction requires a self-consistent solution for the
real space density and velocity fields. The velocity field remains
irrotational when smoothed over large enough scales, so given the
field $\varte$, defined in equation~(\ref{eq:i2}), and appropriate
boundary conditions, the velocity is

\be
\bfv(\bfx) = \f{f(\Omega)}{4\pi} \int {\rm d}^3 x' \varte(\bfx')
\f{\bfx'- \bfx}{\vert \bfx'- \bfx \vert^3} \,.
\label{eq:i4}
\ee 
To proceed further, we need a local estimator of the velocity divergence
from density. Thus, even in density-density comparisons, an inverse
estimator is indispensable. In linear regime $\varte = \de$, hence

\be
\bfv(\bfx) = \f{f(\Omega)}{4\pi} \int {\rm d}^3 x' \de(\bfx')
\f{\bfx'- \bfx}{\vert \bfx'- \bfx \vert^3} \,.
\label{eq:i5}
\ee Yahil \etal (1991) and Strauss \etal (1992) describe an iterative
technique of simultaneously solving for the real space density and
velocity fields, in which they use equation~(\ref{eq:i5}). However, in
the present version of the {\it IRAS}--{\sc potent} comparison, Sigad
\etal (1998) include nonlinear corrections to this equation. The
nonlinear formula for the velocity divergence in terms of the density
they use is a purely phenomenological fit to CDM N-body simulations.
If the ZA is applied to predict density from velocity (the {\sc
potent} reconstruction), why not to apply it to the inverse case as
well, i.e. to predict velocity from density (the {\it IRAS}
reconstruction)?  The reason why Sigad \etal do not do this is simply
that thus far, nobody has succeeded in inverting the ZA. For example,
Nusser \etal (1991) tried to invert it, but failed.

In the present paper I express density in the ZA as a function of the
velocity scalars: the expansion (divergence) and the shear. This
enables me to derive easily the `forward' DVDR in the ZA, i.e., an
analog of formula~(\ref{eq:i3}). This also helps me to invert the ZA,
i.e., to compute the mean velocity divergence given the density
contrast in the ZA (`inverse' DVDR). Such an estimator of the velocity
divergence from density has a scatter, but I show the scatter to be
inevitable if the estimator is to be local. Moreover, I explicitly
compute the scatter and find it to be small. In a follow-up paper,
we test both perturbative and derived here, ZA-based DVDRs against
N-body simulations (Chodorowski \etal 1998b).

The paper is organized as follows: in Section~\ref{sec:vel_scal} I
express the density in the ZA as a local function of the velocity
scalars. In Section~\ref{sec:dens-div} I average this expression to
obtain the mean density given the velocity divergence. In
Section~\ref{sec:scatter} I compute the expected scatter in this
forward DVDR. In Section~\ref{sec:vel_from_dens} I derive an
inverse DVDR, i.e., the mean velocity divergence directly in terms
of the density. Summary and conclusions are given in
Section~\ref{sec:summary}.

\section{Density in terms of the velocity scalars}
\label{sec:vel_scal}

In a Lagrangian approach to PT (Mout\-arde \etal 1991, Bouchet \etal
1992, Bouchet \etal 1995), instead of expanding the density contrast,
one expands the trajectory of a particle,
\be
\bfx = \bfq + D \bfpsi^{(1)}(\bfq) + D^2 \bfpsi^{(2)}(\bfq) + \ldots \,.
\label{eq:n1}
\ee 
Here, $\bfq$ is particle's unperturbed Lagrangian coordinate, $\bfx$
is its final (Eulerian) position, $D(t)$ is the linear growth-factor
of density fluctuations, and $\bfpsi^{(1)}(\bfq)$ and
$\bfpsi^{(2)}(\bfq)$ are the corresponding values of the displacement
fields $\bfpsi^{(1)}$ and $\bfpsi^{(2)}$. The point of the ZA is that
only the field $\bfpsi^{(1)}$ is retained. Then, the velocity of a
particle at an Eulerian position $\bfx$ is simply

\be
\bfv = f(\Omega) H \bfpsi^{(1)}(\bfq) \,,
\label{eq:n2}
\ee
where $H$ is the Hubble constant, and, expressing distances in units
of \kms, in Eulerian space equation~(\ref{eq:n1}) takes the form
\be
\bfq(\bfx) = \bfx - f^{-1} \bfv(\bfx) \,.
\label{eq:n3}
\ee
Hence, the continuity equation reads (Nusser \etal 1991)
\be
\de_{}(\bfx) = || \partial \bfq / \partial \bfx || - 1 = 
|| {\bf I} - f^{-1} \partial \bfv / \partial \bfx || - 1 \,,
\label{eq-1}
\ee
where the double vertical bars denote the determinant and ${\bf I}$ is
the unit matrix.  Expanding the determinant in powers of products of
velocity derivatives we have (following the notation of Sigad \etal
1998)

\be
\de_{}(\bfx) = - f^{-1} \bmath{\nabla} \cdot \bfv + 
f^{-2} \Delta_2 + f^{-3} \Delta_3 \,,
\label{eq-2}
\ee
where

\be
\Delta_2(\bfx) = \sum_{i<j}^3 \left( v_{i,i} v_{j,j} -
v_{i,j}^2 \right)
\label{eq-3} \,,
\ee

\be
\Delta_3(\bfx) = \sum_{i \ne j \ne k}^3 \left( v_{i,i}
v_{j,k} v_{k,j}
- v_{1,i} v_{2,j} v_{3,k} \right)
\label{eq-4} \,,
\ee
and $ v_{i,j} \equiv \partial v_i / \partial x_j$.

In this paper, by the `density in the ZA' I always mean the continuity
density. It is different from the dynamical density in the ZA (i.e.,
resulting from the equation of motion), since the ZA conserves mass
and momentum simultaneously only to first order. The dynamical
density, $\de_d$, is exactly as in linear theory, $\de_d = - f^{-1}
\bmath{\nabla} \cdot \bfv$. Moscardini \etal (1996) used the dynamical
density (specifically, the resulting solution for the velocity in
terms of the ZA-predicted density) to model the velocity field of
clusters of galaxies. A large radius, $20$ \hmpc, of a Gaussian window
with which they smoothed the velocity field makes this approximation
indeed applicable. Density--velocity comparisons, however, employ
considerably smaller smoothing lengths, where the linear DVDR is no
longer valid. As already stated, one approach to find a mildly
nonlinear extension of the linear DVDR is to rigorously derive
higher-order perturbative corrections. A complementary, less rigorous,
but more intuitive and also promising approach is offered by the ZA,
since the {\em continuity\/} density in the ZA successfully recovers
the true density in N-body simulations by Mancinelli \etal (1994) and
Ganon \etal (1998).

Before shell crossing, the velocity field remains irrotational. This
implies that the velocity deformation tensor is symmetric and we can
decompose it into expansion, $\te$, and shear (the traceless part),
$\s_{ij}$:
\be v_{i,j} = {\textstyle \f{1}{3} } \te \de_{ij} +
\s_{ij} \,,
\label{eq-5}
\ee
where in general
\be
\s_{ij} \equiv {\textstyle \f{1}{2} } \left( v_{i,j} +
v_{j,i} \right) - {\textstyle \f{1}{3} } \te \de_{ij}
\label{eq-6}
\ee
and
\be
\te \equiv \bmath{\nabla} \cdot \bfv = v_{k,k}
\,.
\label{eq-7}
\ee 
Here, the symbol $\de_{ij}$ denotes the Kronecker's delta. Note that
$\te = - f \varte$, $\varte$ being defined by equation~(\ref{eq:i2}).
I will now use decomposition~(\ref{eq-5}) in
expressions~(\ref{eq-2})--(\ref{eq-4}), using the methods and the
results of Chodorowski~(1997; hereafter C97). For irrotational fields,
the quantity $\Delta_2$ equals to the quantity $m_{\bfv}$ introduced
by Gramann~(1993). (Note that an expression for $\Delta_2$ in Sigad
\etal 1998 has the wrong sign.) C97 showed that $m_{\bfv} = \te^2/3 -
\s^2/2$ (eq.[37] of C97 with the vorticity term equal to zero), where
$\s^2$ is the shear scalar,

\be
\s^2 = \s_{ij} \s_{ij}
\label{eq-8}
\ee
and I use Einstein's summation convention. Hence, we
have
\be
\Delta_2 = {\textstyle \f{1}{3} }
\left(\te^2 - {\textstyle \f{3}{2} } \s^2 \right)
\,.
\label{eq-9}
\ee
Substitution of decomposition~(\ref{eq-5}) in
equation~(\ref{eq-4})
yields
\begin{eqnarray}
\Delta_3 = \!\!\!\!\!
&-&\!\!\!\!\! || \s_{ij} || \nonumber \\
&+&\!\!\!\!\! {\textstyle \f{1}{3} }
\left(\s_{12}^{2} + \s_{13}^{2} 
+  \s_{23}^{2} - \s_{11} \s_{22} - \s_{11} \s_{33} - \s_{22} \s_{33}
\right) \te \nonumber \\
&-&\!\!\!\!\! {\textstyle \f{1}{27} } \te^3
\,,
\label{eq-10}
\end{eqnarray}
where I have used the property $\s_{ii} = 0$. By
definition,
\be
\s^2 = 2 \left( \s_{12}^{2} + \s_{13}^{2} +  \s_{23}^{2}
\right) +  \s_{11}^{2} + \s_{22}^{2} +  \s_{33}^{2}
\,.
\label{eq-11}
\ee
Using the above equation and the identity
\be
\left( \s_{11} + \s_{22} +  \s_{33} \right)^2 = 0 \,,
\label{eq-12}
\ee
we can cast the second term in equation~(\ref{eq-10}) to
the form
$\s^2 \te /6$. We thus obtain
\begin{eqnarray}
\de_{}(\bfx) = \!\!\!\!\!
&-& \!\!\!\!\! f^{-1} \te + {\textstyle \f{1}{3} } f^{-
2} \left(\te^2 - {\textstyle \f{3}{2} } \s^2 \right) 
\nonumber \\
&+&\!\!\!\!\! 
f^{-3} \left( - || \s_{ij} || + {\textstyle \f{1}{6} }
\s^2 \te
- {\textstyle \f{1}{27} } \te^3 \right)
\,,
\label{eq-13}
\end{eqnarray}
or, using the variables $\varte$ and
\be
\Sig_{ij} \equiv - f^{-1} \s_{ij}
\,,
\label{eq-14}
\ee

\be
\de_{}(\bfx) = \varte + {\textstyle \f{1}{3} } \!
\left(\varte^2 - {\textstyle \f{3}{2} } \Sig^2 \right) +
|| \Sig_{ij} || - {\textstyle \f{1}{6} } \Sig^2 \varte
+ {\textstyle \f{1}{27} } \varte^3
\,,
\label{eq-15}
\ee
where
\be
\Sig^2 = \Sig_{ij} \Sig_{ij} \,.
\label{eq-16}
\ee
Thus, the density contrast is a function of three scalars, constructed
from the velocity deformation tensor: the expansion scalar (the
velocity divergence), the shear scalar, and the determinant of the
shear matrix. The above equation is our starting point to derive the
DVDR within the ZA.

\section{Density in terms of the velocity divergence}
\label{sec:dens-div}

The ZA yields expression~(\ref{eq-2}) for the density in terms of the velocity
derivatives. The resulting expression for the mean density in terms of the
velocity divergence obtains by averaging both sides of equation~(\ref{eq-2})
given the velocity divergence. Having transformed this equation to the
form~(\ref{eq-15}), the conditional averaging is straightforward. We have
\begin{eqnarray}
\lan \de_{} \ran|_{\varte} 
\!\!\!\!\! &=& \!\!\!\!\! \varte +
{\textstyle \f{1}{3} }\! \left(\varte^2 - {\textstyle \f{3}{2} } \lan \Sig^2
\ran|_{\varte} \right) + \lan || \Sig_{ij} || \ran|_{\varte}
- {\textstyle \f{1}{6} } \lan \Sig^2 \ran|_{\varte} \varte
\nonumber \\
\!\!\!\!\! &~& \!\!\!\!\! + {\textstyle \f{1}{27} } \varte^3 
\,.
\label{eq-17}
\end{eqnarray}
The Fourier transform of a shear component is $\left(\Sig_{ij} \right)_{\bfk}
= \left( \hat{k_i} \hat{k_j} - \f{1}{3} \de_{ij} \right) \te_{\bfk}$,
where $\hat{k_i} \equiv k_i/k$ and $\te_{\bfk}$ is the Fourier transform of
the velocity divergence field. Hence,
\begin{eqnarray}
\lan \varte \Sig_{ij} \ran
&=& \int \f{{\rm d}^3 k}{(2 \pi)^3} \,
\left( \hat{k_i} \hat{k_j} - \f{1}{3} \de_{ij} \right)
P_\varte(k)
\nonumber \\
&=& 0
\label{eq-18}
\end{eqnarray}
($P_\varte(k)$ is the power spectrum of the velocity divergence
field). This means that the shear components are uncorrelated with the
velocity divergence. In the case of Gaussian random variables, and
only in this case, it is a sufficient condition to be statistically
independent. Since the initial conditions are assumed to be Gaussian,
in linear regime $\varte$ and $\Sig_{ij}$ are independent. When the
fields become non-linear, they become non-Gaussian as well (e.g.,
Bernardeau \etal 1995, {\L}okas \etal 1995). In the ZA, however, since
the velocity field is proportional to the initial displacement field,
equation~(\ref{eq:n2}), it remains linear even when the density field
becomes non-linear. Derivatives of a Gaussian field, being its linear
combinations, are also Gaussian, so $\varte$ and $\Sig_{ij}$ remain
Gaussian, thus independent. In effect, we can simply replace the
conditional averages in equation~(\ref{eq-17}) by the ordinary
averages. We have $\lan \Sig^2 \ran = (2/3) \veps_\varte^2$. The mean
value of the determinant of the shear matrix is zero. Hence, 

\be
\lan \de_{} \ran|_{\varte} =
a_1^{(ZA)} \varte + a_2^{(ZA)} \left( \varte^2 - \veps_\varte^2 \right) +
a_3^{(ZA)} \varte^3 \,,
\label{eq-19}
\ee
where

\be
a_1^{(ZA)} = 1 - {\textstyle \f{1}{9} } \veps_\varte^2 \,,
\label{eq-20}
\ee
\be
a_2^{(ZA)} = {\textstyle \f{1}{3} }  \,,
\label{eq-21}
\ee
and
\be
a_3^{(ZA)} = {\textstyle \f{1}{27} } \,.
\label{eq-22}
\ee

The mean density given the velocity divergence in the ZA is thus a
third order polynomial in the velocity divergence, similarly to the
third-order PT result~(\ref{eq:i2}). Also the coefficients of the
polynomial are in many aspects similar to the corresponding
coefficients resulting from perturbative calculations: they form a
hierarchy $a_3^{(ZA)} \ll a_2^{(ZA)} \ll a_1^{(ZA)}$, they are
independent of $\Omega$, and $a_1^{(ZA)}$ has a corrective
term,\footnote{ The value of $a_1^{(ZA)}$ is at the leading order
correctly unity, because in the limit of small fluctuations the ZA
recovers linear theory.} which scales linearly with the variance of
the velocity divergence field. (The corrective term is due to the term
$- {\textstyle \f{1}{6} } \Sig^2 \varte$ in equation~[\ref{eq-15}], a
third-order mixed term in the shear and the velocity divergence.)
Quantitatively, however, the coefficients are different. As stated
earlier, in a separate paper we use N-body simulations to test
relative accuracy of both approximations (Chodorowski \etal 1998b).

\section{Scatter in the relation}
\label{sec:scatter}

Expression~(\ref{eq-19}) for density in terms of the velocity
divergence, since obtained by conditional averaging of
equation~(\ref{eq-15}), has clearly a scatter. The rms value of the
scatter at a given value of the velocity divergence, $s|_\varte$, is
given by the square root of the conditional variance, $s|_\varte =
\left.\left\lan \left( \de - \lan \de \ran|_{\varte} \right)^2
\right\ran\right|_{\varte}^{1/2}$.  We have

\be
\left.\left\lan \left( \de - \lan \de \ran|_{\varte} \right)^2
\right\ran\right|_{\varte} = \left.\left\lan \left( 
{\textstyle \f{1}{2}} y + {\textstyle \f{1}{6}} y \varte - ||
\Sig_{ij} || \right)^2 \right\ran\right|_{\varte} \,,
\label{eq-23}
\ee
where 
\be
y \equiv \Sig^2 - \lan \Sig^2 \ran \,.
\label{eq-24}
\ee 
For `typical' fluctuations, the first term in parentheses in
equation~(\ref{eq-23}) is of the order of $\veps_\varte^2$, while the
second and the third are $\calO(\veps_\varte^3)$. In large N-body 
simulations, however, one can trace statistical events of the velocity
field which are many standard deviations away from the mean. In
particular, Chodorowski \etal (1998b) reliably estimate the scatter as
a function of $\varte$ even for $\varte$ well above unity. Therefore,
in equation~(\ref{eq-23}) I do not assume $\varte$ to be small. Since
$\varte$ and $\Sig$ are statistically independent, we obtain

\be
\left.\left\lan \left( \de - \lan \de \ran|_{\varte} \right)^2
\right\ran\right|_{\varte} = {\textstyle \f{1}{4}} \lan y^2 \ran 
\left( 1 + {\textstyle\f{1}{3}} \varte \right)^2 + \calO(\veps_\varte^6)
\,.
\label{eq-25}
\ee
I recall that given Gaussian initial conditions, the velocity field
in the ZA remains Gaussian. For such a field, 
\be 
\lan y^2 \ran = \lan (\Sig^2 - \lan \Sig^2 \ran)^2 \ran = 
{\textstyle \f{8}{45}} \veps_\varte^4
\label{eq-26}
\ee
(see C97 for details), hence

\be
s|_\varte^2 = {\textstyle \f{2}{45}} \veps_\varte^4 
\left( 1 + {\textstyle\f{1}{3}} \varte \right)^2 + \calO(\veps_\varte^6)
\,.
\label{eq-27}
\ee
The probability distribution function for the velocity divergence has
an abrupt cutoff at $\varte = -1.5$, as PT predicts (Bernardeau 1994)
and N-body simulations confirm (Bernardeau \& van de Weygaert 1996,
Chodorowski \& Stompor 1998). Therefore, $1 + \f{1}{3} \varte$ is
always positive and we obtain finally

\be
s|_\varte = b_0^{(ZA)} \veps_\varte^2 + b_1^{(ZA)}
\veps_\varte^2 \varte 
\,,
\label{eq-28}
\ee
where
\be
b_0^{(ZA)} = {\textstyle\f{1}{3}}\sqrt{{\textstyle \f{2}{5}}} 
\simeq 0.21
\label{eq-29}
\ee
and
\be
b_1^{(ZA)} = {\textstyle\f{1}{9}} \sqrt{{\textstyle \f{2}{5}}}
\simeq 0.07
\,.
\label{eq-30}
\ee

Carrying calculations up to second order in PT, C97 derived a formula
for a scatter in the DVDR similar to the first term in
equation~(\ref{eq-28}). Extending the calculations up to third order,
C{\L}PN derived a formula already containing the
second term, but were unable to predict the value of the coefficient
$b_1$. The ZA predicts, in a simple way, not only
formula~(\ref{eq-28}), but the values of both coefficients $b_0$ and
$b_1$ as well.

The rms value of the scatter relative to the rms value of the
divergence, $\veps_\varte$, vanishes in the limit $\veps_\varte \to
0$, as expected. More importantly, however, this ratio is
substantially smaller than unity even for $\veps_\varte$ close to
unity. Thus, even when almost fully nonlinear, the density and the
velocity divergence at a given point remain strongly correlated. In a
follow-up paper we test the prediction of the ZA for a scatter in the
DVDR against N-body simulations.

\section{Velocity from density}
\label{sec:vel_from_dens}

Equation~(\ref{eq-15}) can be perturbatively inverted to express the
velocity divergence as a local function of the density and the
shear. The resulting expansion for $\varte$ has an infinite number of
terms. Up to cubic terms, it is

\be 
\varte(\bfx) = \de - {\textstyle \f{1}{3} } \! \left(\de^2
- {\textstyle \f{3}{2} } \Sig^2 \right) - || \Sig_{ij} || -
{\textstyle \f{1}{6} } \Sig^2 \de + {\textstyle \f{5}{27} } \de^3 +
\calO(\veps_\de^4) , \;
\label{eq-31}
\ee
where $\veps_\de^2 \equiv \lan \de^2 \ran$. Obviously, the velocity
divergence is not a function of the density alone. Thus, like the
forward relation studied in Section~(\ref{sec:dens-div}), a local
estimator of the velocity divergence from the density will inevitably
have a scatter. To obtain an expression for the divergence exclusively
in terms of the density, I will average the above equation {\em given}
the density. We have
\begin{eqnarray}
\lan \varte \ran|_\de 
\!\!\!\!\! &=& \!\!\!\!\! \de - {\textstyle \f{1}{3} } \de^2
+ {\textstyle \f{1}{2} } \lan \Sig^2 \ran|_\de 
- \lan || \Sig_{ij} || \ran|_\de 
- {\textstyle \f{1}{6} } \lan \Sig^2 \ran|_\de \de  
+ {\textstyle \f{5}{27} } \de^3 
\nonumber \\ 
\!\!\!\!\! &~& \!\!\!\!\! + \calO(\veps_\de^4) \,.
\label{eq-32}
\end{eqnarray}
Unlike $\varte$ and $\Sig$, $\de$ and $\Sig$ are {\em not}
independent, since the evolved density is a (mildly) non-Gaussian
variable. The calculation of $\lan \Sig^2 \ran|_\de$ is a non-trivial
problem; I present it in Appendix~A. The result is

\be
\lan \Sig^2 \ran|_\de = {\textstyle \f{2}{3} } \veps_\de^2 
- {\textstyle \f{4}{45} }\veps_\de^2 \de +
\calO(\veps_\de^4) \,.
\label{eq-33}
\ee
Thus, the mean value of the shear scalar given the density weakly
depends on the density. The term generating the dependence is of the
order of $\veps_\de^3$, higher than the constant term. This is because
at the linear order $\de$ and $\Sig$ are independent. For the same
reason, $ \lan || \Sig_{ij} || \ran|_\de = \lan || \Sig_{ij} || \ran +
\calO(\veps_\de^4) = \calO(\veps_\de^4)$. Using this fact and
substituting equation~(\ref{eq-33}) in equation~(\ref{eq-32}) yields

\be
\lan \varte_{} \ran|_{\de} =
r_1^{(ZA)} \de + r_2^{(ZA)} \left( \de^2 - \veps_\de^2 \right) 
+ r_3^{(ZA)} \de^3 
+ \calO(\veps_\de^4) \,,
\label{eq-34}
\ee
where
\be
r_1^{(ZA)} = 1 - {\textstyle \f{7}{45} } \veps_\de^2
\,,
\label{eq-35}
\ee
\be
r_2^{(ZA)} = - {\textstyle \f{1}{3} }
\label{eq-36}
\ee
and
\be
r_3^{(ZA)} = {\textstyle \f{5}{27} } 
\,.
\label{eq-37}
\ee

And if we instead invert formula~(\ref{eq-19}), which expresses the
mean value of density directly in terms of the velocity divergence?
Straightforward inversion of~(\ref{eq-19}) yields
expansion~(\ref{eq-34}), with the coefficients $n_j^{(ZA)}$ which I
will call the na\"{\i}ve ones,

\be
n_1^{(ZA)} = 2 - a_1 - 2 a_2^2 \veps_\de^2 = 
1 - {\textstyle \f{1}{9} } \veps_\de^2
\,,
\label{eq-38}
\ee
\be
n_2^{(ZA)} = - a_2 = - {\textstyle \f{1}{3} }
\label{eq-39}
\ee
and
\be
n_3^{(ZA)} = - a_3 + 2 a_2^2 = {\textstyle \f{5}{27} }
\,.
\label{eq-40}
\ee
The coefficients $n_2^{(ZA)}$ and $n_3^{(ZA)}$ are equal to
$r_2^{(ZA)}$ and $r_3^{(ZA)}$, respectively, but $n_1^{(ZA)}$ is
different from $r_1^{(ZA)}$. This is a consequence of a scatter in the
DVDR. If a relation between two random variables has a scatter, in
general the inverse relation is not given by a mathematical inversion
of the forward relation (e.g., forward and inverse Tully-Fisher
relations). C{\L}PN showed the true and the na\"{\i}ve coefficients to
be related in the following way:

\be
r_1 = n_1 + (2 b_2^2 - b_0^2) \veps_\de^2
\,,
\label{eq-41}
\ee
\be
r_2 = n_2
\label{eq-42}
\ee
and
\be
r_3 = n_3 - b_2^2
\,.
\label{eq-43}
\ee
Here, $b_0$ and $b_2$ are the coefficients entering the {\em
leading-order} perturbative formula for the rms value of the scatter
in the DVDR,

\be
s|_\varte = b_0 \veps_\varte^2 
\left[1 + b_2^2 \varte^2 / (b_0^2 \veps_\varte^2) \right]^{1/2}
+ \calO(\veps_\varte^3)
\label{eq-44}
\ee
(C{\L}PN; note a slightly different notation used here). This
expression was derived under an assumption that $\varte$ is of the
order of $\veps_\varte$, so the second term under square root is of
the order of unity. The formula does not account for the second term
in expression~(\ref{eq-28}) because it is already of third order in
$\veps_\varte$. Comparing expressions~(\ref{eq-44}) and~(\ref{eq-28})
we find that in the ZA the coefficient

\be
b_2^{(ZA)} = 0
\label{eq-45}
\ee

and $b_0^{(ZA)}$ is given by equation~(\ref{eq-29}). From
equations (\ref{eq-42}), (\ref{eq-43}) and~(\ref{eq-45}) we have that
indeed $r_2^{(ZA)} = n_2^{(ZA)}$ and $r_3^{(ZA)} = n_3^{(ZA)}$. Using
equations~(\ref{eq-41}), (\ref{eq-38}), (\ref{eq-29})
and~(\ref{eq-45}) we obtain $r_1^{(ZA)} = 1 - {\textstyle \f{7}{45}}
\veps_\de^2$, in agreement with equation~(\ref{eq-35}). Thus, I
rederived the values of the coefficients $r_j^{(ZA)}$ in a different
way.

Both `forward' and `inverse' relations, expressions~(\ref{eq-19})
and~(\ref{eq-34}), describe mean statistical properties of the matter
field. They were derived by constrained averaging over all possible
realizations of the density and velocity fields. It is instructive to
compare them to the results obtained assuming spherical symmetry of
perturbations. In this case, all the shear terms in
equation~(\ref{eq-15}) vanish, and it simplifies to the form

\be
\de = \left( 1 + \varte/3 \right)^3 - 1 
\,.
\label{eq-46}
\ee
This equation is easily invertible,

\be
\varte = 3 \left[(1 + \de)^{1/3} - 1 \right]  
\,.
\label{eq-47}
\ee 

The above expression is in agreement with a result of Bouchet \etal
(1995) for a spherical top-hat (eq.~[A31] of Bouchet \etal 1995).
When expanded, it yields the values of the coefficients $r_2$ and
$r_3$ equal to these given by equations~(\ref{eq-36})
and~(\ref{eq-37}). It does not, however, predict the correction to the
leading-order value of the linear coefficient $r_1$. More importantly,
it does not involve a constant term. Such a term, $- r_2^{(ZA)}
\veps_\de^2$, is present in equation~(\ref{eq-34}) and naturally
assures the (ordinary) mean of the velocity divergence to vanish, the
property which expression~(\ref{eq-47}) lacks. The shear terms
are thus generally important.

\section{Summary}
\label{sec:summary}
I have derived the mildly nonlinear DVDR as predicted by the ZA, as
well as a scatter in it. The `forward' relation states that the mean
density contrast, given the velocity divergence, is a third-order
polynomial in the velocity divergence. This is `a law of Nature' in
the ZA, or, rather, `a law of the ZA'. In contrast, the `inverse'
relation, expressing the mean velocity divergence in terms of the
density contrast, has an infinite number of terms; I have explicitly
computed the coefficients of the first three. A relation between two
mildly nonlinear variables should be described by a polynomial of
third degree quite well. In any case, I will not attempt to calculate
higher-order coefficients before testing the already computed ones
against N-body simulations: the ZA is only an approximation, and
modelling its prediction even more accurately is no guarantee of
bringing us any closer to the truth.

The $\Omega$-dependence of the DVDR in the ZA enters only via a factor
$f(\Omega)$, used in the definition~(\ref{eq:i2}) of the {\em
scaled\/} velocity divergence. Similarly, in PT, the relation between
the density and the scaled velocity divergence is practically
$\Omega$-independent (Bernardeau~1992, Gramann~1993, C{\L}, C{\L}PN;
cf.\ also Nusser \& Colberg~1998).

I have explicitly computed a scatter in the forward relation; a
scatter in the inverse relation can be computed analogously. I have
computed only the `forward' scatter because here we are mostly
interested in the mean relations and the scatter is only an auxiliary
quantity informing us about the limitations of our local
estimators. The predicted scatter is relatively small, even for the
fields which are almost fully nonlinear. Therefore, the inverse
relation, when obtained by a straightforward inversion of the
forward, will be only slightly biased. Indeed, a proper calculation
of the inverse relation yields a minor correction to the value of
the linear coefficient and no corrections to the quadratic and cubic
coefficients. This offers an efficient way of deriving approximate
values of the coefficients of the higher-order terms, if there is in
future any need to include them.

I have not included the effects of smoothing the evolved density and
velocity fields, while the fields inferred from observations are
smoothed. In rigorous PT, smoothing slightly changes the values of the
`forward' and `inverse' coefficients, making them weakly dependent on
the underlying power spectrum of mass fluctuations and the window
function used (C{\L}, C{\L}PN, Chodorowski \etal 1998b). The inclusion
of smoothing in the ZA can be done in an analogous way to that in
PT. Formula~(\ref{eq-2}), which is clearly for unsmoothed fields, is
however found a successful estimator of smoothed density from smoothed
velocity in N-body simulations by Mancinelli \etal (1994) and Ganon
\etal (1998). Thus, the apparent success of the ZA is somewhat
accidental: neither higher orders in Lagrangian PT nor the smoothing
effects are included, but it still works. If it really works, an
inverse estimator based on the ZA may, on scales smaller than about
$5$ \hmpc, do better than the corresponding estimator based on
third-order (Eulerian) PT. If so, it would be very useful in
large-scale velocity--velocity and density--density comparisons. In a
follow-up paper, we will test the performance of both approximations
in N-body simulations (Chodorowski \etal 1998b).

The perturbative derivation of the mildly nonlinear DVDR by C{\L} and
C{\L}PN is very formal. In contrast, a local relation between the
density and the velocity scalars in the ZA (though strictly valid for
unsmoothed fields only) enables one to derive the DVDR in an easy and
intuitive way. In this picture, the source of the scatter in the DVDR
is the second, `hidden', parameter of the velocity field, namely the
shear. (In reality, another source of the scatter is smoothing.) A
similar conclusion was drawn by C97 on the basis of second-order
PT. However, in the ZA-formula for the density there are terms up to
third order in the velocity derivatives. The third-order terms give
rise to the second term of formula~(\ref{eq-28}) for the scatter,
unpredictable by second-order formalism. Furthermore, a third-order,
mixed term in the shear and the velocity divergence is a source of the
correction to the leading-order value, unity, of the linear
coefficient, $a_1$, in the forward relation. The conditional average
of this term, given the velocity divergence, yields an additional term
linear in the velocity divergence, with the coefficient proportional
to the variance of the velocity divergence field. The linear scaling
of the correction to $a_1$ with the variance is indeed formally
predicted by PT. Thus, even if the ZA fails to model quantitatively
the mildly nonlinear DVDR, it will remain a useful tool to {\em
understand\/} it.

\section*{Acknowledgments}
I wish to thank Stefano Borgani, Adi Nusser and Andy Taylor for
calling my attention to the Zel'dovich approximation in the context of
cosmic density--velocity relations. Francis Bernardeau and Radek
Stompor gave useful comments on an earlier draft of this paper. This
research has been supported in part by the Polish State Committee for
Scientific Research grants No.~2.P03D.008.13 and 2.P03D.004.13, and
the Jumelage program `Astronomie France--Pologne' of CNRS/PAN.

\appendix
\section{Conditional average of the shear scalar}
\label{app:shear}
I outline here a derivation of the expectation value of the velocity
shear scalar {\em given} the velocity divergence or the density
contrast.  The only assumption made is that the density and the
velocity fields remain in the mildly non-linear regime, so the
calculation can be performed perturbatively. Besides that the derived
formula is entirely general, i.e. it can be applied to any
approximation of mildly nonlinear dynamics. Here, I apply it to the
ZA.

The derivation of $\lan \Sig^2 \ran|_\varte$ is greatly simplified
when we introduce an auxiliary variable 

\be
\beta \equiv \varte + \Sig^2 - \lan \Sig^2 \ran 
\,,
\label{eq:a1}
\ee 
where $\lan \Sig^2 \ran$ is an ordinary average of the shear
scalar. Expanding shear components in a perturbative series,
$\Sig_{ij} = \Sig_{ij}^{(1)} + \Sig_{ij}^{(2)} + \ldots$, yields
$\Sig^2 - \lan \Sig^2 \ran = {\Sig^{(1)}}^2 - \lan {\Sig^{(1)}}^2 \ran
+ 2 \Sig_{ij}^{(1)} \Sig_{ij}^{(2)} + \calO(\veps_\varte^4)$, where
${\Sig^{(1)}}^2 \equiv \Sig_{ij}^{(1)} \Sig_{ij}^{(1)}$. Similarly,
$\varte = \varte^{(1)} + \varte^{(2)} + \varte^{(3)} +
\calO(\veps_\varte^4)$, or, for short, $\varte = \varte_1 + \varte_2 +
\varte_3 + \calO(\veps_\varte^4)$. Hence,

\be
\beta = \beta_1 + \beta_2 + \beta_3 + \calO(\veps_\varte^4) 
\,,
\label{eq:a2}
\ee 
where
\be 
\beta_1 = \varte_1
\,,
\label{eq:a3}
\ee
\be
\beta_2 = \varte_2 + \Sig_2, \qquad 
\Sig_2 \equiv {\Sig^{(1)}}^2 - \left\lan {\Sig^{(1)}}^2 \right\ran
\label{eq:a4}
\ee
and
\be
\beta_3 = \varte_3 + \Sig_3, \qquad 
\Sig_3 \equiv 2 \Sig_{ij}^{(1)} \Sig_{ij}^{(2)}
\,.
\label{eq:a5}
\ee

By definition~(\ref{eq:a1}), 

\be
\lan \Sig^2 \ran|_\varte = \lan \Sig^2 \ran + \lan \beta \ran|_\varte
- \varte
\,,
\label{eq:a6}
\ee
so our problem reduces to the calculation of mean $\beta$ given
$\varte$, where both variables, of vanishing mean, are mildly
nonlinear and equal to each other at linear order. This problem was
solved by C{\L}. According to C{\L},

\be
\lan \beta \ran|_{\varte} =
c_1 \varte + c_2 (\varte^2 - \veps_\varte^2) + c_3 \varte^3 + 
\calO(\veps_\varte^4)
\,,
\label{eq:a7}
\ee
where
\begin{eqnarray}
c_1
&=& 1 + \left[ \calZ_2 + \f{(S_{3\beta} - S_{3\varte})\, S_{3\varte} }{3} -
\f{\calZ_4}{2} \right] \veps_\varte^2 \,,
\label{eq:a9} \\
c_2
&=& \f{S_{3\beta} - S_{3\varte}}{6} \,,
\label{eq:a10} \\
c_3
&=& \f{\calZ_4 - (S_{3\beta} - S_{3\varte})\, S_{3\varte} }{6} \,.
\label{eq:a11}
\end{eqnarray}
(I use here slightly different notation.) The quantity $S_{3\beta}$
is defined by

\be
\veps^4 S_{3\beta} = 3 \lan \beta_1^2 \beta_2 \ran
\,,
\label{eq:a12}
\ee
and $S_{3\varte}$ is defined in an analogous way. The quantities
$\calZ_2$ and $\calZ_4$ are given by

\be 
\veps^{4} \calZ_2 = \lan \beta_{2} \varte_{2} \ran - \lan
\varte_{2}^{2} \ran + \lan \beta_{1} \beta_{3} \ran - \lan
\varte_{1} \varte_{3} \ran
\label{eq:a13}
\ee
and
\be
\veps^{6} \calZ_{4} = 3 \lan \beta_{1}^{2} \beta_{2}
\varte_{2} \ran_c - 3 \lan \varte_{1}^{2} \varte_{2}^{2} \ran_c + \lan
\beta_{1}^{3} \beta_{3} \ran_c - \lan \varte_{1}^{3} \varte_{3} \ran_c
\,.
\label{eq:a14}
\ee
In the expressions above, $\veps^2$ is the {\em linear} variance of the
velocity divergence field, $\veps^2 = \lan \varte_1^2 \ran$, and the
symbol $\lan \cdot \ran_c$ stands for the connected (reduced) part of
the moments.

From equations~(\ref{eq:a6}) and~(\ref{eq:a7}) we have 

\be
\lan \Sig^2 \ran|_\varte = \lan \Sig^2 \ran + 
(c_1 - 1) \varte + c_2 (\varte^2 - \veps_\varte^2) + c_3 \varte^3 
\,.
\label{eq:a8}
\ee
Using expansion~(\ref{eq:a2}) of $\beta$ in the expression for the
coefficient $c_2$ yields
\begin{eqnarray}
2 \veps^4 c_2 
&=& \lan \beta_1^2 \beta_2 \ran - \lan \varte_1^2 \varte_2 \ran
\nonumber \\
&=& \lan \varte_1^2 (\beta_2 - \varte_2) \ran \nonumber \\
&=& \left\lan {\varte^{(1)}}^2 \left({\Sig^{(1)}}^2 - 
\left\lan {\Sig^{(1)}}^2 \right\ran \right) \right\ran \nonumber \\ 
&=& \left\lan {\varte^{(1)}}^2 \right\ran 
\left\lan {\Sig^{(1)}}^2 - \left\lan
{\Sig^{(1)}}^2 \right\ran \right\ran \nonumber \\
&=& 0 \,.
\label{eq:a15}
\end{eqnarray}
In the last but one step I used the fact that $\varte$ and $\Sig$ are
independent at linear order (see
Section~\ref{sec:dens-div}). Casting similarly the coefficients $c_1$
and $c_3$ we obtain

\be
\lan \Sig^2 \ran|_\varte = \lan \Sig^2 \ran + 
s_1 \veps_\varte^2 \varte + s_3 \varte^3 + \calO(\veps_\varte^4)
\,,
\label{eq:a16}
\ee
where

\be
s_1 = \calZ_2 - \f{1}{2} \calZ_4
\,,
\label{eq:a17}
\ee
\be
s_3 = \f{1}{6} \calZ_4
\,,
\label{eq:a18}
\ee
with

\be 
\veps^{4} \calZ_{2} = 
\lan \varte_2 \Sig_2 \ran + \lan \varte_1 \Sig_3 \ran 
\label{eq:a19}
\ee
and
\be
\veps^{6} \calZ_{4} = 3 \lan \varte_{1}^{2} \varte_{2}
\Sig_{2} \ran_c + \lan \varte_{1}^{3} \Sig_{3} \ran_c 
\,.
\label{eq:a20}
\ee
Thus, the average value of the shear scalar given the velocity
divergence is equal to its unconstrained average plus the corrective
terms, dependent on the divergence. These terms are of the order of
$\veps_\varte^3$, higher than the first term, which is of the order of
$\veps_\varte^2$. This is again due to the fact that at linear order,
$\Sig$ and $\varte$ are statistically independent.

Though the variable $\varte$ denotes the velocity divergence, the only
specific property of $\varte$ I have used thus far was its
independency of $\Sig$ at linear order. This property is also shared
by the variable $\delta$, since $\delta_1 = \varte_1$. Therefore, an
expression for the mean value of the shear scalar given the density
contrast can immediately be written by replacing the symbol $\varte$
with $\de$ in expression~(\ref{eq:a16}). Specifically,

\be
\lan \Sig^2 \ran|_\de = \lan \Sig^2 \ran + 
s_1' \veps_\de^2 \de + s_3' \de^3 + \calO(\veps_\de^4)
\,,
\label{eq:a21}
\ee
where

\be
s_1' = \calZ_2' - \f{1}{2} \calZ_4'
\,,
\label{eq:a22}
\ee
\be
s_3' = \f{1}{6} \calZ_4'
\,,
\label{eq:a23}
\ee
with

\be 
\veps^{4} \calZ_{2}' = 
\lan \de_2 \Sig_2 \ran + \lan \de_1 \Sig_3 \ran 
\label{eq:a24}
\ee
and
\be
\veps^{6} \calZ_{4}' = 3 \lan \de_{1}^{2} \de_{2}
\Sig_{2} \ran_c + \lan \de_{1}^{3} \Sig_{3} \ran_c 
\,.
\label{eq:a25}
\ee
The formulas~(\ref{eq:a16}) and~(\ref{eq:a21}) are general in a sense
that they are applicable to any approximation of mildly nonlinear
dynamics (including rigorous PT). Here, I will apply
them to the ZA. 

In the ZA, the velocity field remains linear all the time, so
$\varte_2^{(ZA)} = \Sig_3^{(ZA)} = 0$. (The quantity $\Sig_2^{(ZA)}$
is non-zero because it is constructed from first-order
quantities). Hence, $\calZ_2^{(ZA)} = \calZ_4^{(ZA)} = 0$ and

\be
\lan \Sig^2 \ran|_\varte^{(ZA)} = \lan \Sig^2 \ran 
\,.
\label{eq:a26}
\ee
This result is otherwise obvious, since in the ZA, $\Sig$ and $\varte$
remain independent in the nonlinear regime (see
Section~\ref{sec:dens-div}). 

The case of $\lan \Sig^2 \ran|_\de^{(ZA)}$, however, is not so
trivial, because unlike the velocity field, the density field in the
ZA is non-linear. From equation~(\ref{eq-15}),

\be
\de_2^{(ZA)} = {\textstyle \f{1}{3} }
\left({\varte^{(1)}}^2 - {\textstyle \f{3}{2} } {\Sig^{(1)}}^2 \right)
\,.
\label{eq:a27}
\ee
This yields

\be
{\calZ_2'}^{(ZA)} = - \f{\veps^{-4}}{2} \, 
\mathrm{var} \! \left({\Sig^{(1)}}^2 \right)  
\,,
\label{eq:a28}
\ee
where $\mathrm{var} \! \left( {\Sig^{(1)}}^2 \right)$ is the variance
of the linear shear scalar. This variance is given by
equation~(\ref{eq-26}), hence

\be
{\calZ_2'}^{(ZA)} = - \f{4}{45} 
\,.
\label{eq:a29}
\ee
Analogous calculation shows that 

\be
{\calZ_4'}^{(ZA)} = 0 
\label{eq:a30}
\ee
(I recall that $\calZ_4'$ is constructed from the {\em connected}
part of moments). This yields ${s_1'}^{(ZA)} = - 4 /45$ and 
${s_3'}^{(ZA)} = 0$, hence

\be
\lan \Sig^2 \ran|_\de^{(ZA)} = 
\lan \Sig^2 \ran - \f{4}{45} \veps_\de^2 \de + \calO(\veps_\de^4)
\,.
\label{eq:a31}
\ee 
The ordinary average of the shear scalar is equal to $(2/3)
\veps_\varte^2$ and the variance of the velocity divergence field is
not equal to the variance of the nonlinear density field,
$\veps_\de^2$. The difference, however, is
$\calO(\veps_\de^4)$. Therefore finally

\be
\lan \Sig^2 \ran|_\de^{(ZA)} = 
\f{2}{3} \veps_\de^2 - \f{4}{45} \veps_\de^2 \de + \calO(\veps_\de^4)
\,.
\label{eq:a32}
\ee

\end{document}